# Innovations in the field of on-board scheduling technologies

Temenuzhka Avramova, Riccardo Maderna, Alessandro Benetton, Christian Cardenio
AIKO s.r.l.: Autonomous Space Missions
Corso Castelfidardo 30/A, 10129, Torino, Italy
temenuzhka@aikospace.com

**ABSTRACT**

Space missions are characterized by long distances, difficult or unavailable communication and high operating costs. Moreover, complexity has been constantly increasing in recent years. For this reason, improving the autonomy of space operators is an attractive goal to increase the mission reward with lower costs. This paper proposes an onboard scheduler, that integrates inside an onboard software framework for mission autonomy. Given a set of activities, it is responsible for determining the starting time of each activity according to their priority, order constraints, and resource consumption. The presented scheduler is based on linear integer programming and relies on the use of a branch-and-cut solver. The technology has been tested on an Earth Observation scenario, comparing its performance against the state-of-the-art scheduling technology.

## INTRODUCTION

The panorama of space missions has been evolving rapidly in recent years, with increasingly more complex space segment architectures and operations concepts[1]. In parallel, it has been demonstrated that adopting autonomy functionalities on-board leads to a long-term reduction in the costs related to the management of operations, and to additional capabilities being enabled for the space segment. In the frame of advanced automation, several functionalities are essential onboard to enable E4 autonomy level, i.e., "the execution of goal-oriented mission operations on-board" as described by the current standards[2]. Among these, recent innovations from academia and the industry have explored the use of Machine Learning, advanced planning and scheduling, and AI-based failure monitoring on-board.

This paper presents an online, onboard scheduler technology that is fitted to solve the most common problems that are found when managing a space mission, in scenarios such as multi-payload management, resources management, and maneuver planning. The problem approached was characterized by a set of different operational activities that must be scheduled according to several onboard limited resources of different types. Therefore, our objective was to develop a scheduler that arranges these tasks along a timeline in an optimal way according to the constraints imposed by tasks' mutual temporal relations, causal relations, and, obviously, according to the aforementioned set of available resources.

In a simple scenario, the activities to be scheduled can be provided by a ground control center[3, 4]. However, the best performance is obtained when the scheduler is integrated into an environment that enables fully autonomous operations for space missions. In that case, future goals and the associated set of activities are determined by an onboard reasoning module of the satellite. Specifically, the proposed algorithm is constructed to integrate with the MiRAGE library, developed by AIKO[5], which is a software library that employs state of the art technologies in Deep Learning, Expert Systems, and Intelligent Agents to process spacecraft data (telemetry, payload) and take decisions autonomously during the mission. Differently from other autonomy frameworks[6], it does not rely on extensive ground planning but aims to bring the entire decision process onboard. In this regard, we implemented an onboard scheduler with real-time execution, able to react to changes in the scenario due to events detected in orbit and to generate a new timeline for the mission. The development of such an onboard real-time scheduler represents a fundamental step towards fully automated space missions.

In the remainder of the paper, we first define the onboard scheduler, highlighting its the broad modelling capabilities, which cover a wide variety of mission scenarios, task constraints and resources. Then, we present an example of application in the context of an Earth Observation mission. Finally, we compare the performance of the proposed algorithm to those obtained by state-of-the-art scheduling technology, represented by the job scheduler that relies on the CP-SAT solver of Google OR-Tools[7].

## ONBOARD SCHEDULER

The scheduler proposed in this paper takes as input a set T of activities (also called tasks) and is in charge of evaluating, for each task, whether it is possible to schedule the activity and which is its starting time considering priorities, relative order, and consumption of resources. In doing so, the scheduler optimizes the starting time of each task and tries to schedule the greatest number of tasks. The scheduler works with discrete time intervals over a finite horizon H, so that each starting time is an integer value in the interval [0, H]. Therefore, the optimization is expressed as an integer linear programming problem that can be efficiently solved with a branch-and-cut solver, such as the CBC solver[8] used in our implementation.

*Scheduling problem definition*

Each task $T_i \in T$ to be scheduled is defined as a tuple $T_i = (s_i, l_i, schedule\_cost_i, delay\_cost_i)$, where $s_i \in [0, H]$ is the starting time of the task $T_i$ and constitutes a decision variable for the scheduler; $l_i > 0$ is the length (or duration) of the task, so that the task finishes at the time instant $s_i + l_i$. The parameters *schedule_cost* and *delay_cost* are associated with two different aspects of the concept of priority among activities. Specifically, *schedule_cost* is an integer value that goes from 0 to 5, where 0 is a special value that indicates a mandatory task, while the other values define the cost paid in case the task is not inserted in the schedule. If a mandatory activity cannot be inserted in the plan, the scheduler outputs a failure. Instead, $delay\_cost \in [1, 5]$ is directly proportional to the importance to start the task as soon as possible. The sum of all costs associated with the tasks in T defines the cost function of the optimization problem.

The scheduling problem considers several types of constraints. First, the execution of each task might be bounded inside a specific time interval $[a_i, b_i]$, named *feasible_interval*, which enforces the constraint $s_i \in [a_i, b_i - l_i]$ on the starting time. This constraint models activities such as downlink operations of a satellite that could be scheduled only during the visibility window with the ground station.

Furthermore, each task can occupy or consume some resources. Three categories of resources are considered to cover most of the cases that are found in practice:

- *binary resources;*
- *multiple resources;*
- *consumable resources.*

Binary resources cannot be used by more than one task at the same time and lead to non-overlapping constraints among activities. More formally, if $b_i(t)$ indicates the number of tasks that uses the i-th binary resource at the unit time t, then it must hold that:

$$b_i(t) \leq 1, \forall t \in [0, H] \qquad (1)$$

An example of a binary resource is the usage of science equipment, such as the camera of a satellite: if two tasks require access to the camera, they cannot be scheduled to run in parallel.

The multiple resources category contains resources that are not consumed but have a maximum capacity that limits simultaneous use. For example, it is possible to have an upper limit of the power consumption per unit of time. In this case, all tasks will be scheduled according to their consumption of power and groups of tasks that overcome the threshold of simultaneous consumption, will not be overlapped. More formally, if $m_i(t)$ indicates the total usage of the i-th multiple resource at the unit time t, that is the sum of the amounts consumed by each task executing at time t, and $M_i$ indicates the total availability of that resource per unit time, then it must hold that:

$$m_i(t) \leq M_i, \forall t \in [0, H] \qquad (2)$$

The resources in the last category, called consumable resources, are the one that can be consumed and/or generated in time. An immediate example is the consumption of propellant or battery storage. Also in this case, each task is associated with the amount of resource it consumes (or generates). For example, a task that acquires pictures of the Earth consumes electricity, while a maintenance task like battery charging, produces it. Giving the sequence of tasks, the scheduler takes into account that at any time the availability of all consumable resources must be non-negative. More formally, if $c_i(t)$ indicates the cumulative consumption of the i-th consumable resource at the unit time t, that means the sum of all the consumption (or generation) of that resource over the time interval [0, t], and if $C_i$ is its initial availability, then it must hold that:

$$c_i(t) \leq C_i, \forall t \in [0, H] \qquad (3)$$

The final set of constraints is constituted by precedence constraints. In situations where more tasks are needed to achieve a goal, the execution of some activities must likely respect a default order. For instance, the goal of image acquisition could be composed of the following tasks: "point camera", "prepare camera", "acquire images" and "save images". In this case, it is apparent that the ordering among the tasks is essential. The developed scheduler considers two types of precedence constraints:

- *tight* precedence, that is the second task $T_j$ must start *exactly* after the preceding one $T_i$ plus an offset dt:
  $s_i + l_i = s_j - dt$

- *lax* precedence, that is the second task $T_j$ can start at any moment after the end of the preceding one $T_i$ plus an offset dt:
  $s_i + l_i \leq s_j - dt$

The offset value can be either positive or negative, so that it is possible to model both the case when the second task must start after the first one has finished and the one when the second task starts while the first is still executing. Tight constraints are used to describe precondition relations among tasks, i.e., cases when the preceding task is a precondition for the execution of the following task that cannot be executed otherwise (e.g., pointing the camera toward the target before acquisition). Instead, lax constraints express logic precedence resulting from goal-related reasoning that impose the desired ordering among specific tasks. Nevertheless, the execution of each task is independent. For instance, one may want to perform a downlink operation only after an acquisition operation, since otherwise there would be no data to downlink.

At this point, an observation must be done that stems from the introduction of precedence constraints. Some tasks might not be scheduled because of incompatibility in the constraints or insufficient availability of resources (the parameter *schedule_cost* is the one that penalizes the missed schedule of a task). If this happens to a task that participates in a precedence constraint, two cases can occur that are supported by the scheduler. First, if the first task in the constraint is not useful on its own but only as a precondition for the execution of the next one, then the entire sequence loses its meaning. Therefore, either both or none of the tasks must be inserted into the schedule. Conversely, the second scenario is when it is also meaningful to schedule only the preceding task in the constraint, while the following can be excluded from the schedule if needed. Overall, the scheduler implements four types of precedence constraints: *tight-both*, *tight-at-least-the-first*, *lax-both*, *lax-at-least-the-first*.

### Online rescheduling

Once that the scheduler solves the optimization problem and finds a suitable schedule for the input set of activities and constraints, the plan is dispatched to a lower-level execution layer of the spacecraft. However, the occurrence of unexpected events during operation, such as faults or unforeseen environment conditions, or the release of a new goal can invalidate the current schedule and call for a rescheduling procedure[9]. In this case, a new set of activities is provided as input that must be co-scheduled together with those already present in the schedule and not already executed.

However, particular care must be placed on the rescheduling of the tasks that are running when rescheduling is triggered. Indeed, each task has additional information on whether it allows or forbids preemption. Specifically, we identify four behaviors for running tasks:

- *No preemption*: the task cannot be interrupted and will be concluded in the new schedule;
- *From start*: the task can be interrupted and must be completely rescheduled in another time of the horizon;
- *From interrupt*: the task can be interrupted and only the remaining part must be rescheduled in another time of the horizon;
- *Delete*: the task cannot be rescheduled and is deleted from the schedule.

When an activity that belongs to a precedence sequence is preempted and rescheduled, then also all the other tasks in the chain, that is all the tasks involved in *tight* precedence constraints with the interrupted task, have to be rescheduled, even those that have already been executed. This is clearer if we refer to the example of image acquisition: suppose that the "acquire images" task is rescheduled after other activities, then the spacecraft must execute again the "point camera" and "prepare camera" activities to perform the target task correctly.

A final case to be considered is when the arrival of a new goal (and associated activities) makes some of the tasks that were already in the schedule obsolete. For example, if the satellite is acquiring images with a high-frequency camera and some clouds are detected, then a new goal can be set to acquire images with the low-frequency camera to not waste energy and memory space. In this case, it is apparent that rescheduling both the tasks related to high-frequency and low-frequency acquisition is useless. This situation is handled by associate additional information to incoming tasks about which other tasks, possibly present among the pending ones, must be discarded. When a task is removed from the scheduling problem, all other tasks that follow the deleted task inside a tight constraint are also removed from the pending activities. Instead, tasks that follow a lax constraint are maintained, but the constraint is removed.

**Table 1: Tasks to schedule with parameters and indication of resource consumption/generation**

| Task | l | delay _cost | schedule _cost | Preemption | feasible_ interval | Resources (units per time) | | |
|---|---|---|---|---|---|---|---|---|
| | | | | | | Binary | Multiple | Consumable |
| AcquisitionHF | 20 | 4 | 3 | Interrupt | [] | Camera Position | Power (40) | Battery (3) |
| BatteryCharge | 10 | 3 | 3 | No | [0,15] | Position | | Battery (-10) |
| DownlinkSat2Ground | 10 | 2 | 3 | No | [5,45] | Antenna Position | Power (40) | Battery (3) |
| DownlinkSat2Sat | 5 | 5 | 3 | No | [0,10] | Antenna Position | Power (40) | Battery (3) |
| SetHFOption | 1 | 3 | 3 | Interrupt | [] | Camera | Power (20) | Battery (1) |
| SetPositionK | 1 | 3 | 3 | No | [] | Reaction Wheel Position | Power (20) | Battery (1) Propellant (2) |
| SetPositionX | 1 | 3 | 3 | No | [] | Reaction Wheel Position | Power (20) | Battery (1) Propellant (2) |
| SetPositionY | 1 | 3 | 3 | No | [] | Reaction Wheel Position | Power (20) | Battery (1) Propellant (2) |
| SetPositionZ | 1 | 3 | 3 | No | [] | Reaction Wheel Position | Power (20) | Battery (1) Propellant (2) |

**EXAMPLE SCENARIO**

This section illustrates a realistic scenario that covers some of the cases presented in the previous section as an example. We consider an Earth Observation application where a Low Earth Orbit satellite aims to acquire high-quality images of the Earth's surface. To do so, it can perform two different types of image acquisition activities, one that acquires images at high frequency (higher memory and energy consumption, but better-quality target observation) and another one that acquires images at low frequency (useful to monitor the target while it is not ready to be observed, for example because covered by clouds). In addition, the satellite can downlink data to ground stations, but also spread messages to other satellites (e.g., for coordination purposes in constellations). The resources that are managed by the scheduler are:

- *Binary resources*: Reaction wheel, Position;
- *Multiple resources*: Power, with a maximum capacity of 60W per unit of time;
- *Consumable resources:* Battery, with initial availability of 20 units, and Propellant, with initial availability of 100 units.

Where the binary resource on the satellite position is introduced to avoid overlapping tasks that need a specific attitude configuration.

Suppose that at a given time the current goals of the satellites are: 1) Acquire images of the Earth's surface, 2) Downlink observation data to a ground station, 3) Send data to another satellite, 4) Recharge batteries (by pointing solar panels toward the sun). Table 1 lists the tasks that must be scheduled to reach all goals together with their associated parameters and information related to the usage of resources for each task. Furthermore, the tasks have the following precedence constraints, where x ≺ y stands for "task x precedes task y":

*tight-both:*   SetPositionX ≺ DownlinkSat2Sat
SetPositionZ ≺ AcquisitionHF
SetPositionK ≺ BatteryCharge
SetPositionY ≺ DownlinkSat2Ground

*lax-both:*   SetHFOption ≺ AcquisitionHF

*lax-at-least-the-first:*
AcquisitionHF ≺ DownlinkSat2Ground

Running the scheduler with the horizon set to H = 50 units of time, the result obtained is the one shown in Figure 1. By analysing the schedule, all precedence and resource constraints are met. Moreover, one can notice that the scheduler tries to plan tasks in parallel when possible, such as with SetHFOption and SetPOsitionX at the beginning. In this case, the two activities do not require common equipment and consume a total of 40W of Power, which is below the maximum usage threshold. Instead, all the other tasks in the example require maintaining a specific satellite configuration during execution, which is modelled as a binary resource and prevents the overlap of the activities.

**Table 2: Additional tasks for rescheduling together with parameters, indication of resource consumption/generation, and of the previous tasks made obsolete by the new one**

| Task | l | delay_cost | schedule_cost | Preemption | feasible_interval | Resources (units per time) | | | Obsolete tasks |
|------|---|------------|---------------|------------|-------------------|---------|----------|------------|----------------|
| | | | | | | Binary | Multiple | Consumable | |
| AcquisitionLF | 15 | 5 | 3 | Interrupt | [] | Position Camera | Power (20) | Battery (1) | AcquisitionHF |
| SetLFOption | 1 | 3 | 3 | Interrupt | [] | Camera | Power (20) | Battery (1) | SetHFOption |
| SetPositionZ | 1 | 3 | 3 | No | [] | Reaction Wheel Position | Power (20) | Battery (1) Propellant (2) | |

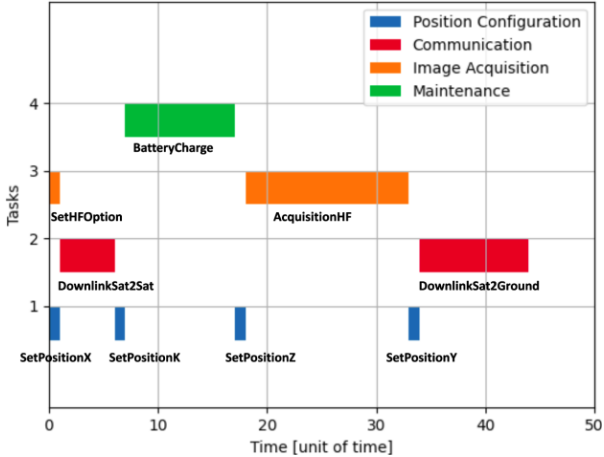

**Figure 1: Schedule obtained with the activities in Tables 1 as input**

Going deeper in the analysis, we can see that the task BatteryCharge is scheduled before AcquisitionHF although the latter has a higher *delay_cost*. Moreover, no precedence constraints require a particular ordering between the two. The reason why the execution of BatteryCharge is anticipated even if it leads to a higher cost schedule is that the scheduling of tasks DownlinkSat2Sat and DowlinkSat2Ground one after the other, without a recharging task in between, would run out the battery storage. Instead, the scheduler can correctly take into account the consumption of consumable resources and plan the proactive behaviour of the satellite.

To test the online rescheduling capabilities, we suppose the arrival of a new goal at time 15. Specifically, we simulate the fact that, thanks to a cloud detection module, such as the one included in the MiRAGE library, the satellite realizes that the observation target is covered and that there is no need to acquire images anymore. Therefore, the onboard reasoning module sends a new goal of monitoring the target via low-frequency imaging. The new input tasks to the scheduling problem and their constraints are detailed in Table 2. In particular, since we are dealing with a rescheduling stage, the table also reports the indication of which tasks from Table 1 are made obsolete by each of the new tasks. Obsolete activities are removed from the pending list and are not included in subsequent schedules. Moreover, the tasks have the following precedence constraints:

*tight-both:*      SetPositionZ ≺ AcquisitionLF

*lax-both:*      SetLFOption ≺ AcquisitionLF

Therefore, a new schedule is generated for the time interval of [15, 65], which is reported in Figure 2. When rescheduling takes place, the task BatteryCharge is running. Since the ongoing task does not allow preemption, it is correctly allocated to continue its execution in the new schedule. Moreover, the new task SetLFOption is planned to run in parallel with BatteryCharge. In fact, the scheduler considers the whole horizon to be available for scheduling any of the pending tasks even when time slots are already populated by activities inherited by the previous iteration.

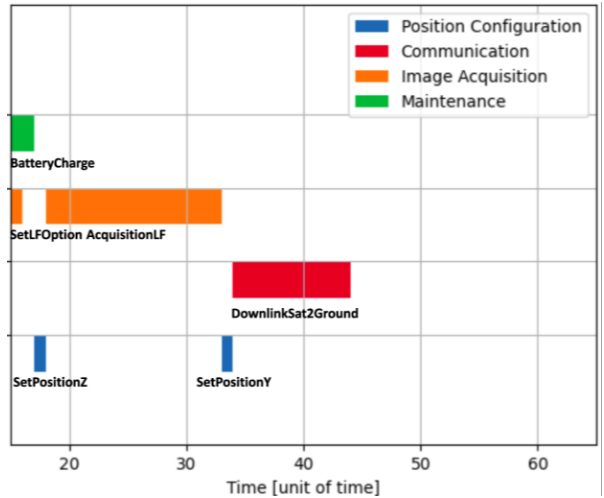

**Figure 2: Schedule obtained after rescheduling with the activities in Table 2 as input**

An important thing to notice is that obsolete tasks related to high-frequency image acquisition

(SetHFOption and AcquisitionHF) are substituted by those related to low-frequency acquisition (SetLFOption and AcquisitionLF). Instead, DownlinkSat2Ground is kept from the previous schedule, as it was not indicated as a task to be deleted by any of the new tasks. This is done regardless of the presence in the first scheduling problem of a lax precedence constraint between AcquisitionHF and DownnlinkSat2Ground. As already mentioned, lax constraints that contain removed tasks are deleted from the scheduling problem, whereas the surviving task can still be scheduled unless explicitly removed from the pending list. This allows the reasoning module of the autonomy software to profit from the visibility window of the ground station to send payload data independently of the presence of the AcquisitionHF task. This is beneficial to free storage space for future acquisitions.

**PERFORMANCE ANALYSIS**

A crucial point, when dealing with onboard real-time scheduling, is the performance in terms of computation time, that is, given a set of tasks to be scheduled, the amount of time needed to provide a schedule to the operation system of a satellite. Since all the different classes of constraints included in the scheduling problem can be expressed as different linear inequalities, the solving time of the scheduler strictly depends not only on the number of tasks and the length of the planning horizon, but also on the type of constraints. To have a point of comparison on the execution time, another scheduler, with the same characteristics as the one presented here, has been constructed using the functions of the Job Problem of Google OR-Tools[7].

A test campaign on Linux desktop environment has been carried out considering a set of problems with a different number of tasks to be scheduled (5, 10, and 15) and, for each case, the inclusion of different combinations of resource constraints (no resource constraints, binary resources only, multiple resources only, consumable resource only). Each scenario has been tested on 7 different problems with a horizon length of 336 time units to average the performance.

In Figure 3, the average execution time is grouped in boxplots by the number of tasks (5, 10, and 15) for both the presented scheduler and the OR-Tools one. From the plots, it is apparent that the proposed scheduler is faster than OR-Tools solution with an average scheduling time of about 0.2, 0.6 and 0.9 seconds for 5, 10 and 15 tasks, respectively. Also, both approaches show linear complexity in the number of activities. Note that the execution time includes both the initialization of the problem and the computation of the schedule. Both stages are important in the presented application, as the scheduler must run in real-time and reschedule online with different tasks and constraints.

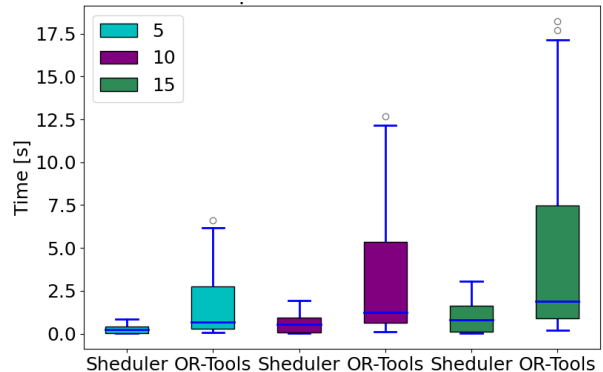

**Figure 3: Execution times of the two schedulers for 5, 10 and 15 tasks (H = 336)**

The substantial difference in performances between the two schedulers shown in Figure 4 derives especially from the poor performance of the OR-Tools algorithm in presence of consumable resources (see Figure 4). The difference in the averages of the execution time between the proposed scheduler and OR-Tools grows with the increasing complexity of the constraints included in the problem. When resource constraints are not considered, OR-Tools performs slightly better than our algorithm (0.6 s against 0.8 s). However, the proposed scheduler does not suffer from the addition of resources and, on the contrary, better exploits the presence of more constraints that reduces the feasible solution space. Instead, Google OR-Tools suffer from the inclusion of more complex constraints with increasing scheduling time required when multiple resources and, especially, consumable resources are present. In particular, the required time soar to about 11 s in the latter case, against 0.3 s required by the proposed scheduler in the same conditions. Table 3 reports the average running time for all considered cases.

**Table 3: Average execution time grouped by the different resource constraints included**

| [s] | No resources | Binary resources | Multiple resources | Consumable resources |
|---|---|---|---|---|
| Proposed | 0.8 | 0.8 | 0.3 | 0.3 |
| OR-Tools | 0.6 | 1.1 | 1.5 | 11.0 |

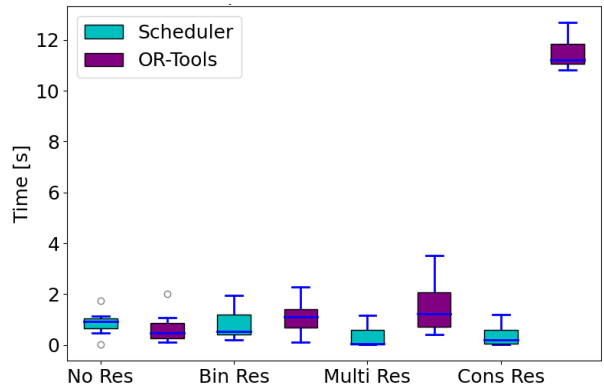

**Figure 4: Execution times of the two schedulers grouped by the different classes of resource constraints in the problem (10 tasks, H = 336)**

By investigating the running time of the OR-Tools version of the scheduler in detail, one can understand that the bottleneck is constituted by the time needed to initialize the problem, that is the time needed to declare variables and constraints, and not by the solving time. Indeed, the implementation of OR-Tools makes the addition of resource constraints heavy from a computational point of view. This is more apparent with consumable resources since their inequalities must keep all the information of the past times and, therefore, they become longer in time.

## CONCLUSIONS

This paper presents an online scheduler that can be integrated inside an onboard autonomy software framework. The scheduler is specifically developed for space missions, such as satellite activities, to enable greater autonomy and take advantage of its benefits. The approach provides wide modelling capabilities to cover all the relevant scenarios and accounts for priorities and precedence among tasks, and different types of resources. Moreover, it is capable of rescheduling in response to changes of goals or unexpected events. Results show that the scheduler can compute optimal plans that satisfy all constraints with a competitive running time, which confirms the capability of the scheduler to be suitable for onboard real-time scheduling.